\documentclass[useAMS,usenatbib]{mn2e}
\twocolumn
\usepackage{bm}
\usepackage{amsfonts}
\usepackage{amsmath}
\usepackage{graphicx}
\usepackage{subfigure}
\usepackage{times}

\def\gsim{\;\lower4pt\hbox{${\buildrel\displaystyle >\over\sim}$}\;}
\def\lsim{\;\lower4pt\hbox{${\buildrel\displaystyle <\over\sim}$}\;}
\def\grls{\;\lower4pt\hbox{${\buildrel\displaystyle >\over <}$}\;}
\title[Jupiter Bursty Radio Emissions Observed at Urumqi]
{Bursty Variations of Jovian 6cm Radio Emissions \\
 and Quasi-Periodic Jupiter's Polar Activities}
\author[Y.-Q. Lou, H. G. Song, Y. Y. Liu, M. Yang]
{Yu-Qing Lou$^{1,\ 2}$, Huagang Song$^{3,\ 4,\ 5}$, Yinyu Liu$^{1}$,
Meng Yang$^{1}$\\
$^1$ Physics Department and Tsinghua Centre for Astrophysics
 (THCA), Tsinghua University, Beijing, 100084, China\\
$^2$ National Astronomical Observatories, Chinese Academy of
 Sciences (CAS), A20, Datun Road, Beijing 100012, China\\
$^3$ Xinjiang Astronomical Observatory, Chinese Academy of
Sciences (CAS), Urumqi, Xinjiang 830011, China\\
$^4$ Graduate University of the Chinese Academy of Sciences (CAS),
Beijing 100049, China\\
$^5$ Key Laboratory of Radio Astronomy, Chinese Academy of
Sciences (CAS), Urumqi, Xinjiang 830011, China}
\date{Accepted 2011 December 12.
    Received 2011 December 8; in original form 2011 April 13}
\pagerange{\pageref{firstpage}--\pageref{lastpage}} \pubyear{2009}

\def\LaTeX{L\kern-.36em\raise.3ex\hbox{a}\kern-.15em
    T\kern-.1667em\lower.7ex\hbox{E}\kern-.125emX}

\begin{document}
\label{firstpage}
\maketitle

\begin{abstract}
In reference to Jupiter south polar quasi-periodic
  40-50 min (QP-40) activities and the model
  scenario for global QP-40 oscillations of the Jovian
  inner radiation belt (IRB), we validate relevant
  predictions and confirmations by amassing empirical
  evidence from Ulysses, Cassini, Chandra, Galileo,
  XMM-Newton, and Advanced Composition Explorer for
  Jupiter north polar QP-40 activities.
We report ground 6cm radio observations of Jupiter by
  Urumqi 25m telescope for synchrotron intensity bursty
  variations of the Jovian IRB and show their likely
  correlations with the recurrent arrival of high-speed
  solar winds at Jupiter.
\end{abstract}


\section{Introduction}

The phenomenology of Jupiter south polar
  QP-40 activities (including QP-40 bursts
  of relativistic electrons and of associated
  low-frequency radio emissions with partially
  {\it right-hand} polarizations)
  discovered by experiments of
  Ulysses spacecraft in 1992-1993 (Simpson et al. 1992;
  MacDowall et al. 1993; McKibben et al.
  1993; Zhang et al. 1995) prompted the theoretical model scenario
  for global QP-40 magneto-inertial oscillations of the Jovian
  inner radiation belt (IRB) (Lou 2001a; Lou \& Zheng 2003 -- LZ
  hereafter) recurrently triggered by the arrival of intermittent
  high-speed solar winds at the Jovian magnetosphere
  (Bame et al. 1992, 1993).
Variations of solar wind ram pressure stir the IRB and
  the Jovian sunward magnetosphere at all radii by
  magnetohydrodynamic (MHD) waves and adjustments,
  irrespective of specific substructures therein
  (e.g. Waite et al. 2001; Samsonov et al. 2011;
  Halekas et al. 2011).
Variations of IRB may be detectable because of its synchrotron
  radio emissions from relativistic electrons trapped inside.
Being an important planetary source of relativistic
  electrons in the global heliosphere, the Jovian IRB
  traps relativistic electrons as well as ions and is
  known to be an intense source of synchrotron radio
  emissions (Ginzburg \& Syrovatskii 1965).
Such global QP-40 magneto-inertial oscillations of Jupiter's
  IRB may trigger QP-40 circumpolar leakage of relativistic
  electrons from irregular cusp zones (sandwiched between closed
  and open magnetic fields) with accompanying low-frequency
  radio bursts almost along local Jovian magnetic field lines.
By this reasoning, the global QP-40 IRB oscillation scenario
  predicts similar Jovian north polar QP-40 activities but
  with partially {\it left-hand} circular polarizations for
  north QP-40 low-frequency radio bursts (Lou 2001a; LZ;
  for circular polarizations of Jovian nKOM, see Kaiser \& Desch 1980)
  as later confirmed by spacecraft observations of Ulysses, Galileo,
  Cassini, Chandra, XMM-Newton and
  ACE during the Ulysses distant
  encounter with Jupiter in 2000-2004 in the north heliosphere
  (Menietti et al. 2001; Bolton et al. 2002; Smith
  et al. 2003; Elsner et al. 2005; Kimura et al. 2008).
The remarkable north (perhaps also south) polar QP-40 X-ray
  brightness pulsations of hot spots
  observed on 18 Dec 2000 (Gladstone et al. 2002; for
  Jovian polar far UV/X-ray auroral activities and pulsed
  reconnections at the dayside magnetopause, see Bunce et al.
  2004) were found to likely correlate with the arrival of
  a high-speed solar wind of $\sim$700$\hbox{ km s}^{-1}$
  at Jupiter by LZ
  in contrast to the two non-pulsating phases of
  Jovian polar X-ray brightness for $\sim$40 hr
  during 24-26 Feb 2003
  (Gurnett et al. 2002; Elsner et al. 2005)
  and for $\sim$60 hr during 25-29 Nov 2003
  (Branduardi-Raymont et al. 2007) with likely
  slower solar wind speeds of $\sim 450$-$500\hbox{ km s}^{-1}$
  at Jupiter estimated in this Letter.
This plausible high-speed solar wind stimulated
  IRB oscillation scenario predicts occasional
  synchrotron IRB bursty pulsations
  which may be detectable by ground radio
  telescopes (Lou 2001a; LZ).
We report here
  6cm Jovian observations by the 25m Urumqi telescope for
  IRB bursty variations possibly stimulated by high-speed solar
  winds at Jupiter using the relevant ACE and ephemeris data.
This would offer a valuable diagnostic prospect to coordinate
  ground and space campaigns in various combinations to
  observe Jovian IRB dynamics and QP-40 polar activities.
In broader contexts of geophysics and astrophysics, this research
  is pertinent to the dynamics and radiation of Earth/planet
  magnetospheres (e.g. Walt 1994) and of pulsar magnetospheres
  (e.g. Lou 2001b, 2002).

\section{Key Facts of Jupiter QP-40 Activities }

On the way to explore
  the global heliosphere at high latitudes, Ulysses
  first reached Jupiter in early Feb 1992.
Approaching Jupiter within its magnetosphere
  and roughly in the ecliptic plane,
  Ulysses detected QP-15 min radio bursts [panel
  (a) in fig. 2 and table 2 of MacDowall et al. 1993].
Such QP-15 bursts at frequencies 1-50 kHz, named ``Jovian
  Type III bursts" by their dynamic spectral nature, are
  pulse-like VLF/LF emissions
  discovered earlier by Voyager (Kurth et al. 1989).
Subsequently outbound from the Jovian south pole at higher
  latitudes $\sim 30-40\hbox{ S}^{\circ}$, Ulysses
  unexpectedly intercepted well correlated QP-40 min series
  bursts of relativistic electrons (with energy $E\gsim 8.9$
  MeV) as well as other energetic ions detected by the
  COSPIN
  (Simpson et al. 1992; McKibben et al.
  1993)
  and of partially {\it right-hand} circular radio emissions
  (frequency $\nu\lsim$0.7 MHz -- mostly in the range of
  $1-200$ kHz) received by the
  URAP
  (Simpson et al. 1992; MacDowall et al. 1993;
  McKibben et al. 1993; Zhang et al. 1995)
  from the south polar direction.
Adjacent bursts are $\sim 40-50$ min apart and each of such
  radio bursts has a typical emitted power of $\sim$$10^8$W
  and lasts $\sim$5$-$10 min [panel (a) in fig. 2 and table 2
  of MacDowall et al. 1993; Menietti et al. 2001].
Travelling from the Jovian magnetosphere towards the Sun in
  the southern heliosphere, Ulysses no longer detected
  QP-40 relativistic electron bursts but continued to receive
  several thousands of such QP-40 radio bursts which bear a
  strong correlation with recurrent high-speed
  ($\gsim 400-500\hbox{ km s}^{-1}$) solar winds at Jupiter
  (figs 11 and 12 of
  MacDowall et al. 1993; Bame et al. 1992, 1993).
The source or acceleration of relativistic electrons and
  QP-40 mechanisms challenge our theoretical and empirical
  understanding of Jovian QP-40 activities first detected
  from Jupiter's south pole.

\section{
IRB QP-40 magneto-inertial oscillations}\label{sec:Model}

With a grossly dipolar magnetic field and polar
  surface field strengths of $\sim$10$-$14.4G
  (e.g. Hess et al. 2011), Jupiter possesses an IRB
  in the radial range of $\sim$1.5$-3.0 R_J$
  that traps relativistic electrons with energies up to
  $\sim$50MeV or higher as revealed by intense IRB
  synchrotron radio emissions in $\sim 2.2-90$cm wavelengths
  (e.g.
  de Pater 1990; Sault et al. 1997; Bolton et al. 2002).
A rotating magnetic dipole of Jupiter's IRB may support
  magneto-inertial oscillations with periods of $\sim 40-50$
  min ($n=0$) and shorter periods for higher harmonics
  (Lou 2001a).
By formula (9) of Lou (2001a) with $n=1, 2, 3, 4$
 therein, the quasi-period ranges are $23-30$,
 $18-22$, $15-19$, $13-17$ min, respectively.
These oscillation modes are hybrid, involving rotation
  and MHD waves in dipolar IRB magnetic fields
  anchored at Jupiter.
Such global IRB pulsations may be triggered by
  drastic compressions of sunward Jovian magnetosphere due
  to the recurrent arrival of fast solar winds;
the rotating Jovian magnetosphere tends to over-corotate
  by angular momentum conservation due to the drastic
  contraction of extended sunward magnetosphere
  and the inevitable wind speed intermittency leads to IRB
  MHD adjustments and thus stimulated IRB oscillations
  (Lou 2001a; see estimates of IRB synchrotron
  emissions in Sec. 3.2 of LZ).
Such global QP-40 IRB oscillations may
 disturb circumpolar and polar magnetospheric plasmas
  (e.g. Balogh et al. 1992) in a QP-40 manner
  in {\it both} polar caps of Jupiter.
As speculated, those circumpolar zones separating the polar
  open field and the IRB closed field may release relativistic
  electrons and ions in irregular segments from the
  IRB in series of QP-40 bursts as modulated by QP-40 IRB
  oscillations during certain phases.
Jupiter is known to be an important planetary source
  of relativistic electrons to populate the entire heliosphere
  (e.g. Nishida 1976;
  Lou 1996).
As the Jovian magnetic field points towards its south pole,
  out-streaming extremely relativistic electrons gyrate
  around south field lines with very small pitch
  angles and emit low-frequency ($\nu\lsim$0.7 MHz) beamed
  radio bursts with partially {\it right-hand}
  circular polarizations
  (MacDowall et al. 1993; Lou 2001a; LZ).
As the Jovian magnetic field
  points outward from north pole, out-streaming extremely
  relativistic electrons gyrate around north circumpolar
  magnetic field lines with very small pitch angles and emit
  low-frequency ($\nu\lsim$0.7 MHz) beamed radio bursts yet
  with partially {\it left-hand} circular polarizations
  (see Kaiser \& Desch 1980 for nKOM).
This IRB oscillation scenario may explain why Jovian QP-40
  activities correlate with recurrent fast solar winds at
  Jupiter (MacDowall et al. 1993; Bame et al. 1993; Lou 2001a; LZ).
It is crucial to collect more observational evidence to
  validate this model scenario in pertinent aspects.

\section{Pertinent Predictions and confirmations}
\label{sec:QP40}

There are consequences for the scenario of global
  QP-40 IRB magneto-inertial oscillations stimulated
  by recurrent high-speed solar winds at Jupiter.
We first anticipate similar QP-40 polar activities to also
  occur around the Jovian north pole (Lou 2001a; LZ).
Except for in situ speed measurements near Jupiter, most
  of our solar wind speeds at Jupiter are inferred from
  ACE and ephemeris data.
We simply assume that a wind speed intercepted by ACE persists.
However, complications of MHD wind interactions and solar
  variations will compromise our wind speed estimates at Jupiter.
Our statements should be all qualified in this regard (see fig. 5
  of Zieger \& Hansen 2008 for uncertainties in wind speed inferences).

For the second ``distant encounter" at a
  distance of $\sim 0.8-2$ AU from Jupiter
  in Feb 2004,
  Ulysses observed Jovian radio waves from
  high to low north latitudes ($+80^{\circ}$ to $+10^{\circ}$
  Jovicentric latitude) for 6 months
  (e.g. Smith et al. 2003; Kimura et al. 2008).
Two types of QP radio bursts were detected, i.e.
  quasi-periods shorter or longer than $\sim 30$ min,
  and they emerged between
  $\sim 0.4-1.4R_J$ above the north pole of Jupiter
  (Kimura et al. 2008).
The Ulysses and ACE data confirmed what we had
  anticipated for fast solar wind stimulated global QP-40
  IRB oscillations to trigger north circumpolar QP-40 bursts
  of relativistic electrons escaping from the synchrotron
  IRB and accompanying low-frequency QP-40 radio bursts
  (Lou 2001a; LZ).
Fig 2 of  
   Kimura et al. (2008) shows a refreshing example of QP-40
   radio bursts on 7 Feb 2004 from the Jovian north
   pole (see fig 2 of
   MacDowall et al. 1993 for comparison);
the relevant ephemeris and ACE data of 22 Jan 2004
   seemingly confirm an arrival of shocked high-speed
   $\sim$$580\hbox{ km s}^{-1}$ solar wind at
   Jupiter on 7 Feb 2004 for $\sim$$6-12$ hr (Fig. 2);
at this shock front, the upstream wind speed
   was close to $\sim$$690\hbox{ km s}^{-1}$.
We shall detail this type of combined ephemeris and solar
   wind data analysis presently for Urumqi 25m telescope
   6cm radio observations of Jupiter.
Another important supporting evidence comes from four series
  of QP-40 radio bursts detected by Ulysses on 6 Oct 2003
  shown in fig 10 of
  Elsner et al. (2005); by checking the relevant ACE data
  of 9 Oct 2003, this case seems to involve a fast
  solar wind of $\sim$$600\hbox{ km s}^{-1}$ at Jupiter
  on 6 Oct 2003 (Fig. 2);
the wind was maintained at $\sim$$550-650$
  km s$^{-1}$ for $\sim$1 day.
Radio observations (10-20 kHz) of Ulysses during $24-26$
  Feb 2003 together with simultaneous Chandra X-ray
  observations show no evidence for strong QP-40 oscillations
  (fig. 10 of Elsner et al. 2005).
The pertinent ephemeris and ACE data analysis indicate a likely
  arrival wind speed of $\lsim$500 km s$^{-1}$ at Jupiter;
  the wind was maintained at $\sim$$430-500$ km s$^{-1}$ for
  $\sim$5 days (Fig. 2).
These cases appear to agree with our anticipation
  (Lou 2001a; LZ) that north QP-40 radio bursts also
  correlate with recurrent fast solar winds at
  Jupiter similar to south polar QP-40 activities.
Within the Jovian magnetosphere and by the analogy
  to what occurred around the south pole, it is
  most likely (though impossible to verify)
  that QP-40 bursts of extremely
  relativistic electrons from Jovian north circumpolar region
  also accompanied these low-frequency QP-40 radio bursts
  detected by Ulysses from the north pole
  (Lou 2001a; LZ).
The Jovian magnetic field points outwards from the north pole.
  QP-40 low-frequency radio bursts beamed from out-streaming
  extremely relativistic electrons gyrating around magnetic
  field lines with tiny pitch angles from north
  circumpolar zones were predicted to give rise
  to partially {\it left-hand} circular polarizations
  (Lou 2001a; LZ) and this prediction appears
  now confirmed by Ulysses/URAP observations
  (Kimura private communications 2010).

\section{QP-40 polar pulsations of X-ray hot spots}\label{sec:Xray}

On 18 Dec 2000 (10$-$20 UT) in support of the Cassini
  fly-by of Jupiter, the Chandra
  HRC targeted Jupiter
  for $\sim$10 hr and revealed remarkable QP-45 pulsations
  of X-ray brightness in the hot spot inside the north
  Jovian auroral oval
  (fig 3 of Gladstone et al. 2002).
For such strikingly similar quasi-periods of $\sim$40$-$45 min,
  this X-ray diagnostics might be supportive to and consistent
  with global Jovian IRB QP-40 polar activities
  (Lou 2001a) and associated polar QP-40 magnetospheric oscillations
  (LZ), and we speculate that QP-40 bursts of relativistic
  electrons and accompanying {\it left-hand} circular low-frequency
  radio emissions might have emerged from north circumpolar zones
  during such polar QP-40 X-ray hot spot pulsations.
In our scenario, QP-40 IRB oscillations are probably stimulated
  by the drastic Jovian magnetospheric compression due to
  high-speed intermittent solar wind; QP-40 IRB magnetic
  field disturbances may affect adjacent polar open
  magnetic fields across the oval zones around the poles
  (Southwood \& Hughes 1983).
Fast MHD waves or magnetosonic modes may effectively transmit
  QP-40 MHD disturbances over magnetic polar caps.
Thus, global QP-40 IRB oscillations also make polar open
  magnetic field to pulsate via transmissions of MHD waves
  across the polar caps by compression and rarefaction of
  polar field lines and plasmas
  (Samsonov et al. 2011; Halekas et al. 2011).
To buttress this hypothesis, we examined the relevant ephemeris
  and ACE data of 9 Dec 2000 (Fig. 2) and verified that a
  persistent $\sim 700\hbox{ km s}^{-1}$ solar wind likely
  arrived Jupiter on 18 Dec 2000 (LZ).
On 24-26 Feb 2003, there was another campaign
  of Chandra, Ulysses radio,
  HST UV targeting Jupiter (Elsner et al. 2005). 
Neither 40 hr Chandra X-ray (both north and south) nor 72 hr
  Ulysses radio (north) data in fig 10 of Elsner et al. (2005)
  show evidence of QP-40 oscillations.
The pertinent ephemeris and ACE data of 10-12 Feb 2003 indicate
  that during $\sim$3 days, Jupiter may have encountered relatively
  slow winds of $\sim$$450-500\hbox{ km s}^{-1}$.
XMM-Newton observed Jupiter from 23:00 25 Nov to 12:00 29 Nov
  2003 for 245 ks and found no $\sim$45 min X-ray
  pulsations (Branduardi-Raymont et al. 2007).
The solar wind speed was likely $\sim$$400-500\hbox{ km s}^{-1}$
  at Jupiter for 25-29 Nov 2003 by the ephemeris and ACE data.
These three contrasting cases of 18 Dec 2000
  (Gladstone et al. 2002; LZ), 24-26 Feb 2003
  (Elsner et al. 2005) and 25-29 Nov 2003
  (Branduardi-Raymont et al. 2007) offer preliminary
  supporting evidence that QP-40 X-ray pulsations in both
  north and south poles might possibly correlate with the
  recurrent arrival of sufficiently fast solar winds at Jupiter.
Meanwhile, the combined analysis of Ulysses and ACE
  data (Fig. 2) on the three cases of 6 Oct 2003,
  7 Feb 2004  
  and 24-26 Feb 2003 
  may suggest that north partially left-hand circular QP-40
  radio bursts likely correlate with the recurrent fast solar
  winds at Jupiter.
By speculative inferences, such partially {\it left-hand} circular
  QP-40 radio bursts from Jupiter north pole should have been
  produced by QP-40 bursts of extremely relativistic electrons
  streaming out of north circumpolar regions.

\section{Urumqi 6cm radio observations of Jupiter}
\label{sec:6cm}

The predicted QP-40 synchrotron IRB variations excited and
  sustained by recurrent fast solar winds at Jupiter may
  be captured by ground radio telescopes (Lou 2001a; LZ).
If firmly established, this would provide observational
  verification and a strong empirical link for the global
  IRB QP-40 oscillation scenario and pertinent phenomena
  of Jovian QP-40 polar activities.
Due to the on-and-off nature of Jovian QP-40 polar activities
  and practical constraints of telescope time allocation, it
  is only by chance to observe Jupiter's IRB bursty variations
  given available times of the 6cm receiver of
  Urumqi 25m radio telescope.
This telescope with a 3m secondary mirror was built
  in 1991-1993, commenced to operate in Oct 1994, and has been
  active in international VLBI collaborations and for the
  Chinese VLBI in the two successful ChangE Moon missions.
The antenna was refurbished and calibrated in May
  2005 for a higher surface accuracy of $\lsim 0.40$ mm (rms).
The telescope works at 1.3, 2.8, 3.6/13, 6, 18/21, 49, and 92cm
  with receivers mounted at either primary or secondary focus.
The antenna sits at longitude 87$^{\circ}$10.67' East and
  latitude 43$^{\circ}$28.27' North with the track level
  at $2080.5$m above sea level.
With the primary focal length of 7m and a pointing accuracy
  $\lsim 15''$ (rms) and for the efficient and convenient
  data reduction, we used the 6cm receiver in both
  the vertical and horizontal scanning modes with a
  resolution (HPBW) of 10.066$^{\prime}$.
The 6cm receiver remains cooled for a central frequency
  at 4620MHz or 6.49cm wavelength with a bandwidth of
  100$-$490 MHz and an antenna efficiency $\sim$52.2\%.

Radio flux densities were determined with ``cross-scans"
 in both azimuth and elevation, four at each point.
This enables us to check the pointing offsets in both coordinates.
A Gaussian fit is performed for every subscan.
 The peak amplitude of such Gaussian fit measures
 the flux density of each pointing.
After applying a correction for small pointing offsets, peak
 amplitudes of both AZ and EL in one cross-scan are averaged.
We next correct the measurements for atmospheric effects and the
 antenna gain (the elevation-dependent effect), using secondary
 flux calibrators which are known to show no variations on short
 timescales, and correct the remaining, systematic time-dependent
 effects in the measured flux densities.
Error bars include subscan errors
 and those of pointing corrections.
Finally, we check our observations against an absolute flux
 density scale using the primary calibrator quasar 3C286.

The distance range between the Earth and Jupiter is
  $\sim$4$-$6 AU yearly.
Jupiter radius $R_J=7.14\times 10^9$ cm
  spans angular sizes of $0.4'$ to $0.27'$.
The radial range of Jovian IRB spans $\sim$$1.5-3.5R_J$
  at 6.1cm
  (fig. 5 of de Pater 1990).
Thus, the beam of 25m telescope completely covers
  Jupiter and IRB together for 6cm emissions.
As Jupiter rotates in $\sim$10 hr period, the mean
  intensity variation can be detected as in Figs. 1,
  3, 4 and \ref{Fig5} (also fig. 1 of de Pater 1990).
Depending on seasonal, weather and local conditions,
  we can track Jupiter for $\sim 5-9$ hrs daily.

On 1 Jan 2009, we observed Jupiter using
  the 6cm receiver of Urumqi 25m telescope.
Shown in Fig. 1, bursty flux intensity variations with
  $\sim$20-40 min timescales (a mean timescale of $\sim$22 min)
We also show the stable radio flux density
  of planetary nebula NGC 7027 as the control calibrator.
Using the pertinent ACE data (see Fig. 2), we show below
  that this epoch may coincide with a rapid rise of
  solar wind speed at Jupiter from $\sim$290$\hbox{ km s}^{-1}$
  up to $\sim$550$\hbox{ km s}^{-1}$ and a sustained solar wind
  speed $\gsim$500$\hbox{ km s}^{-1}$ thereafter for $\sim$ 1 day.
By the ephemeris, the distance from the Sun to Jupiter is
  5.1129AU in Jan 2009.
For a solar wind speed of $\sim$$550\hbox{ km s}^{-1}$, this
  wind stream left the Sun 16.09 days earlier on 
  16 Dec 2008 when the longitude of Jupiter is
  302.32$^{\circ}$ relative to the Equinox.
On 2008-12-16, 
  the Earth is at 0.984 AU from the Sun with a longitude
  84.77$^{\circ}$ relative to the Equinox and an orbital
  speed of 1.08$^{\circ}$ per day.
For such an isolated fast wind stream to persist one to
  several solar rotations as evidenced by the ACE data
  of several months, this wind stream would reach the
  Earth 15.0 days later for a solar rotation period of
  $\sim$$27$ days (e.g. Lou 1987),
  i.e. early on 2008-12-31. 
From the ACE data, the solar wind speed is
  $\sim$$550\hbox{ km s}^{-1}$ on 2008-12-31 noon (Fig. 2). 
The explicit analysis here serves as an example; all
  cases of our solar wind speed estimates at
  Jupiter are performed in the same spirit
  (more documentation of the ACE SWEPAM
  level 2 data can be found at the website http:
  //www.srl.caltech.edu/ACE/ASC/level2/swepam\_l2desc.html).

For the 6cm Jupiter observation at Urumqi
  on 19 Jan 2008 in Fig. 3, we examined the
  relevant ACE data of 20 and 21 Jan 2008.
On 19 Jan 2008, Jupiter encountered a fast
  wind of $\sim$$ 580\hbox{ km s}^{-1}$
  and had a longitude of $273.20^{\circ}$
  relative to the Equinox;
this same wind stream was intercepted by ACE
  between late 20 and before 21 Jan 2008 (Fig. 2).
As this wind was launched from the Sun in early 3 Jan
  2008, the Earth longitude was $\sim$$102.05^{\circ}$
  relative to the Equinox.
The mean intensity of Jupiter was $\sim$$ 4.05$ Jy and there
  were 4 bursts relative to the mean Jovian spin
  variation and the stable control calibrator quasar 3C286.
The 6cm data on 20 Jan 2008 (Fig. 4) also indicated
  bursty features while the flux control calibrator
  quasar 3C286 remained stable.
The ACE data of 21 Jan 2008 indicate that Jupiter on 20
  Jan 2008 in Fig. 2 also encountered a
  $\sim$$ 550\hbox{ km s}^{-1}$ solar wind which was
  later intercepted by ACE early 21 Jan 2008.
This wind speed pattern persisted recurrently and can be
  identified in Dec 2007 and Feb 2008.
%
%

We also observed Jupiter at Urumqi on 5
 and 15
 Dec 2008 for $\sim$6 and $\sim$6.5 hrs,
 with quasar 3C286 as control calibrator
 (see Fig. \ref{Fig5}).
Our 6cm data show slow variations of Jovian 10 hr rotation
 but no QP bursts, and the pertinent ACE data indicate
 arrival wind speeds $\lsim 460$ km s$^{-1}$ at Jupiter
 for these two cases.

\section{Summary and Conclusions}\label{conclusions}

We advanced the model scenario for global QP-40 IRB
  magneto-inertial oscillations excited by the drastic
  compression of sunward Jovian magnetosphere due to
  the recurrent arrival of fast and intermittent
  solar winds (Lou 2001a)
  to first explain south polar QP-40 activities as
  discovered by Ulysses
  (Simpson et al. 1992; MacDowall et al. 1993).
We collect here more observational evidence to support our
  model predictions for north polar QP-40 activities of
  Jupiter combining Ulysses and ACE data
  (Elsner et al. 2005; Kimura et al. 2008).
North Jovian polar QP-45 pulsations of X-ray hot spots
  (Gladstone et al. 2002) might be caused by
  QP-40 polar magnetospheric oscillations associated
  with fast solar wind driven QP-40 IRB oscillations
  (Lou 2001a; LZ).
The two contrasting Chandra cases
  (Gladstone et al. 2002; Elsner et al. 2005) and the
  XMM-Newton 2003 Nov case (Branduardi-Raymont 2007)
  together with the pertinent ACE and ephemeris data analysis here
  might reveal a possible correlation between polar QP-45
  X-ray pulsations and sufficiently fast solar winds at Jupiter (LZ).
In particular, we present preliminary evidence for possible
  bursty variations of the Jovian IRB on timescales of $\sim 20-60$
  min as revealed by direct 6cm radio observations using the Urumqi
  25m telescope in XinJiang;
  again, such IRB activities seem to correlate with the
  recurrent arrival of high-speed solar winds at Jupiter
  according to the pertinent ACE and ephemeris data.
We speculate that such IRB bursty variations excited and sustained
  by the recurrent arrival of high-speed solar winds at Jupiter
  may also manifest at radio wavelengths longer or shorter than
  6cm and be captured by ground radio telescopes under proper
  conditions (Lou 2001a).
We hope to stimulate well-planned concerted campaigns of ground
  and space observations to probe Jovian IRB bursty variations
  in the near future.

\begin{figure}
\begin{center}
    \includegraphics[height=5.0cm,width=7.0cm]{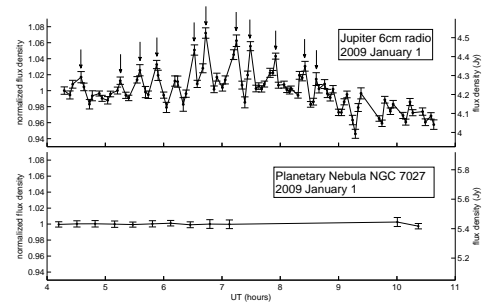}
\end{center}
\caption {
  Observations of Jupiter on 1 Jan 2009
  by the Urumqi 25m radio telescope with the 6cm receiver
  and of the stable control calibrator planetary
  nebula NGC 7027.
The abscissa is
  UT in hrs. Between
  5 and 8 hrs, eight radio flux density peaks emerge,
  giving a mean timescale of $\sim 22$ min.
  }
  \label{Fig1}
\end{figure}

\begin{figure}
\begin{center}
   \includegraphics[height=14cm,width=6.5cm]{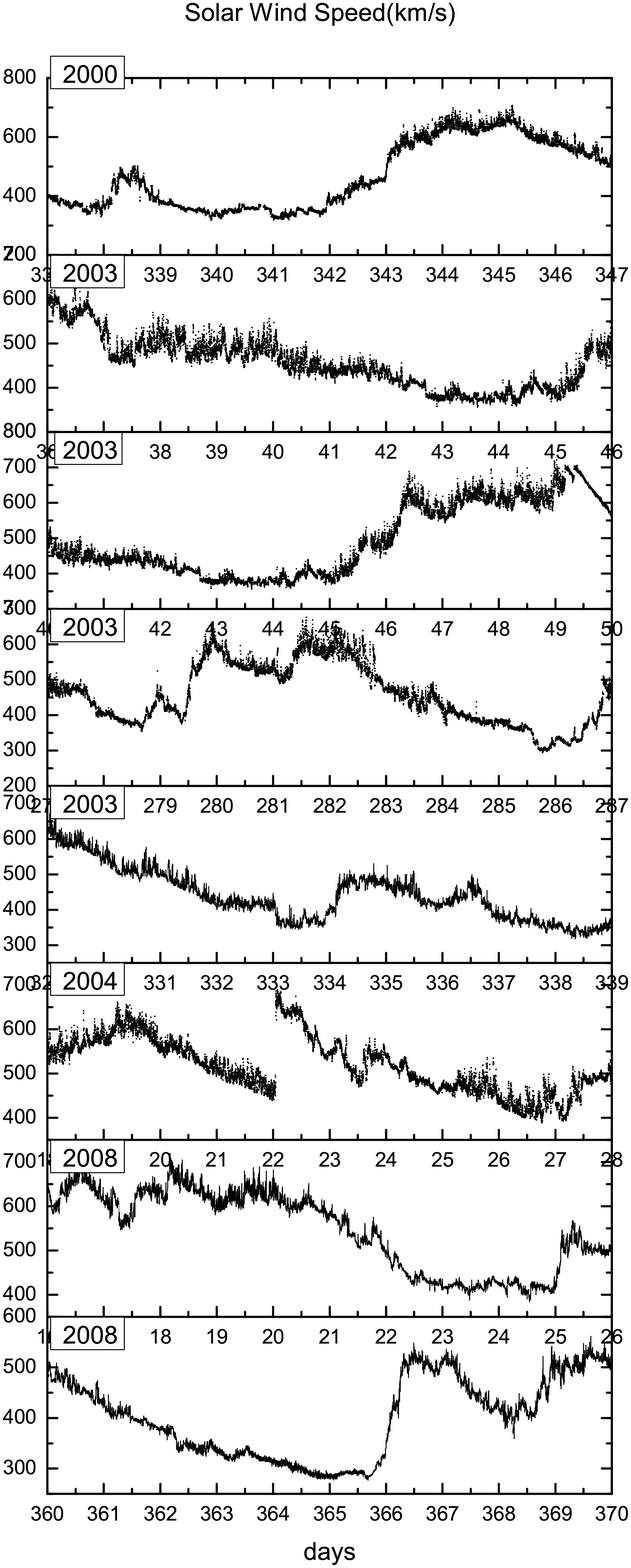}
\end{center}
\caption {
  Pertinent ACE
  solar wind speed data
  for cases of 18 Dec 2000, 24-26
  Feb 2003, 24-26 Feb 2003, 6 Oct 2003, 25-29 Nov 2003, 7 Feb
  2004, 19-20 Jan 2008, and 1 Jan 2009 from top to bottom.
}
  \label{Fig2}
\end{figure}

\begin{figure}
\begin{center}
    \includegraphics[height=4.7cm,width=7.0cm]{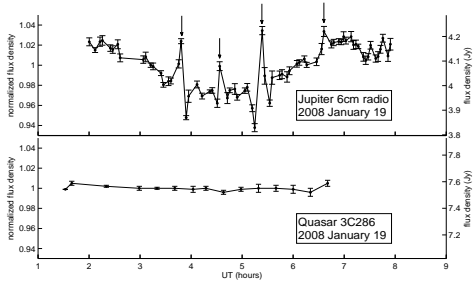}
\end{center}
\caption {
  Observations of Jupiter on 19 Jan 2008
  by the Urumqi
  6cm receiver
  for 4
  major peaks and of the stable quasar
  3C286 as the flux calibrator
  with errors.
The 4 major consecutive peaks are separated
  by $45$, $50$ and $73$ min.
The $\sim$10 hr mean Jovian rotational
  variation can be discerned.
The ephemeris and ACE data for 20 Jan 2008 (Fig. 2)
  indicated a likely sustained high-speed solar wind
  of $\sim$600$\hbox{ km s}^{-1}$ for $\sim 1$ day.
  }
  \label{Fig3}
\end{figure}

\begin{figure}
\begin{center}
  \includegraphics[height=5.0cm,width=7.0cm]{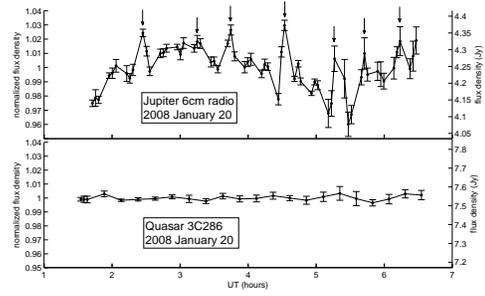}
\end{center}
\caption {
  Observations of Jupiter on 20 Jan 2008
  by Urumqi
  6cm receiver
  with several major peaks and of the stable
  quasar 3C286 as the flux calibrator
  with errors.
The radio peak time separations are 48, 30,
  47, 44, 26, and 32 min (left to right).
By the ephemeris and ACE data of 21 Jan 2008
  (Fig. 2), there was a likely sustained solar wind
  of $\sim 550\hbox{ km s}^{-1}$ for $\sim 1$ day.
  }
  \label{Fig4}
\end{figure}

\begin{figure}
\begin{center}
  \includegraphics[height=3.8cm,width=7.0cm]{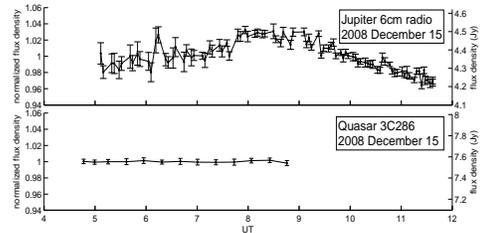}
\end{center}
\caption {Urumqi 6cm observations of Jupiter on 15 Dec 2008
  and of the stable 3C286 as the flux calibrator.
By the ephemeris and ACE data, there was a likely sustained
  solar wind of $\lsim$$440\hbox{ km s}^{-1}$.
The case of 5 Dec 2008 is similar to this case with
  a likely wind speed $\lsim$$460\hbox{ km s}^{-1}$.}
  \label{Fig5}
\end{figure}

\section*{Acknowledgments}

This research was supported in part by
  THCA, by
  NSFC grants 10373009, 10533020, 11073014 and J0630317
  at Tsinghua U., by the Key Laboratory of Radio
  Astronomy in CAS,
  by MOST grant 2012CB821800,
  by Tsinghua University Initiative Scientific Research
  Program, by the Yangtze Endowment, the SRFDP 20050003088,
  200800030071 and 20110002110008,
  and the Special Endowment for Tsinghua College Talent
  Program from MoE.


\label{lastpage}

\begin{thebibliography}{07}

\bibitem[\protect\citeauthoryear{Balogh et al. 1992}{1992}]{Balogh1992}
Balogh A., et al. Science, 257, 1515

\bibitem[\protect\citeauthoryear{Bame1992}{1992}]{Bame1992}
  Bame S. J., et al. 1992, Science, 257, 1539

\bibitem[\protect\citeauthoryear{Bame1993}{1993}]{Bame1993}
  Bame S. J., et al. 1993, Geophys. Res. Lett., 20, 2323


\bibitem[\protect\citeauthoryear{Bolton et al. 2002}{2002}]{Bolton2002}
  Bolton S. J., et al. 2002, Nature, 415, 987

%
%
%
%

\bibitem[\protect\citeauthoryear{Branduardi-Raymont2007}{2007}]
{BranduardiRaymont2007}Branduardi-Raymont G., et al. 2007, A\&A,
463, 761

\bibitem[\protect\citeauthoryear{Bunceetal2004}{2004}]
 {Bunce2004}Bunce E. J., et al.
 2004, JGR, 109, A09S13


\bibitem[\protect\citeauthoryear{de Pater 1990}{1990}]{dePater1990}
de Pater I., 1990, ARA\&A, 28, 347

\bibitem[\protect\citeauthoryear{Elsner et al. 2005}{2005}]{Elsner2005}
  Elsner R. F., et al. 2005, J. Geophys. Res., 110, A01207
%
%
%
%
%
%
%
%
%
%

\bibitem[\protect\citeauthoryear{Ginzburg1965}{1965}]{Ginzburg1965}
Ginzburg V. L., Syrovatskii S. I., 1965, ARA\&A, 3, 297

\bibitem[\protect\citeauthoryear{Gladstone et al. 2002}{2002}]{Gladstone2002}
Gladstone G. R., et al. 2002, Nature, 415, 1000

\bibitem[\protect\citeauthoryear{Gurnett et al. 2002}{2002}]{Gurnett2002}
Gurnett D. A., et al. 2002, Nature, 415, 985



\bibitem[\protect\citeauthoryear{Halekasetal}{2011}]{Halekas2011}
Halekas J. S., et al. 2011, JGR, 116, A07222
%

\bibitem[\protect\citeauthoryear{Hess et al 2011}{2011}]{Hess2011}
 Hess S. L. G., et al.
 2011, JGR, 116, A05217

\bibitem[\protect\citeauthoryear{KaiserDesch1980}{1980}]{KD1980}
 Kaiser M. L., Desch M. D., 1980, GRL, 7, 389
%

%

\bibitem[\protect\citeauthoryear{Kimura et al. 2008b}{2008}]{Kimura2008b}
Kimura T., et al., 2008, Planet. Space Sci., 56, 1967
%
%
%
%
%

\bibitem[\protect\citeauthoryear{Kurth et al. 1989}{1989}]{Kurth1989}
  Kurth W. S., et al.
  1989, J. Geophys. Res., 94, 6917

\bibitem[\protect\citeauthoryear{Lou 1987}{1987}]{Lou1987}
  Lou Y.-Q., 1987, ApJ, 322, 862

\bibitem[\protect\citeauthoryear{Lou 1996}{1996}]{Lou1996}
  Lou Y.-Q., 1996, Geophys. Res. Lett., 23, 609

\bibitem[\protect\citeauthoryear{Lou 2001a}{2001a}]{Lou2001a}
  Lou Y.-Q., 2001a, ApJ, 548, 460

\bibitem[\protect\citeauthoryear{Lou 2001b}{2001b}]{Lou2001b}
  Lou Y.-Q., 2001b, ApJ Lett., 563, L147

\bibitem[\protect\citeauthoryear{Lou 2002}{2002}]{Lou2002}
  Lou Y.-Q., 2002, ApJ Lett., 572, L91

\bibitem[\protect\citeauthoryear{Lou & Zheng 2003}{2003}]{LouZheng2003}
  Lou Y.-Q., Zheng C., 2003, MNRAS Lett., 344, L1 (LZ)

\bibitem[\protect\citeauthoryear{MacDowall1993}{1993}]{MacDowall1993}
  MacDowall R. J., et al. 1993, Planet. Space Sci., 41, 1059

%
%
%
%
%
%
%
%
%
%
%
%
%
%


\bibitem[\protect\citeauthoryear{McKibbenSimpsonZhang1993}{1993}]{McKibben1993}
 McKibben R. B., et al.
 1993, Planet. Space Sci., 41, 1041

\bibitem[\protect\citeauthoryear{Menietti et al. 2001}{2001}]
{Menietti2001} Menietti J. D., et al.
 2001, Radio Sci., 36, 815

\bibitem[\protect\citeauthoryear{Nishida 1976}{1976}]{Nishida1976}
Nishida A., 1976, J. Geophys. Res., 81, 1771


%

\bibitem[\protect\citeauthoryear{Samsonov et al.}{2011}]{Samsonov2011}
Samsonov A. A., et al. 2011, JGR,
116, A10216

\bibitem[\protect\citeauthoryear{Sault et al.}{1997}]{Sault1997}
Sault R. J., et al.
 1997, A\&A, 324, 1190

\bibitem[\protect\citeauthoryear{Simpson et al.}{1992}]{Simpson1992}
Simpson J. A., et al. 1992, Science, 257, 1543

\bibitem[\protect\citeauthoryear{Smith et al.}{2003}]{Smith2003}
Smith E. J., et al. 2003, Science, 302, 1165

\bibitem[\protect\citeauthoryear{SouthwoodHughes}{1983}]{SH1983}
Southwood D. J., Hughes W. J., 1983, SSRev, 35, 301

\bibitem[\protect\citeauthoryear{Stone et al.}{1992}]{Stone1992}
Stone R. G., et al. 1992, Science, 257, 1524

\bibitem[\protect\citeauthoryear{Waite et al.}{2001}]{Waite2001}
Waite J. H., et al. 2001, Nature, 410, 787
%

\bibitem[\protect\citeauthoryear{Walt 1994}
{1994}]{Walt}Walt M., 1994, Introduction to Geomagnetically
Trapped Radiation (Cambridge: Cambridge University Press)


\bibitem[\protect\citeauthoryear{Zhang et al.}{1995}]{Zhang1995}
Zhang M., et al. 1995, J. Geophys. Res., 100, 19497

\bibitem[\protect\citeauthoryear{Zieger & Hansen}{2008}]
{ZiegerHansen2008} Zieger B., Hansen K. C., 2008, J.
 Geophys. Res., 113, A08107

\end{thebibliography}
\end{document}